\documentstyle[epsf,11pt]{aipproc}
\begin{document}
\begin{flushright}
UH511-996-02 \\
March 2002
\end{flushright}
\title{Constraining four neutrino mass patterns from neutrinoless double
beta decay}
\author{Sandip Pakvasa\footnote{Email:  pakvasa@phys.hawaii.edu} and Probir Roy\footnote{Permanent address:  Tata Institute of Fundamental Research, Homi
Bhabha Road, Mumbai 400 005, India;  Email: probir@phys.hawaii.edu}}
\vspace{.75 in}
\address{\large{Department of Physics and Astronomy, \\
University of Hawaii, \\
Honolulu, HI  96822, USA}}
\maketitle
\vspace{.75 in}

\begin{center}
Abstract
\end{center}
All existing data on neutrino oscillations (including those from the LSND
experiment) imply a four neutrino scheme with six
different allowed mass patterns.  Some of the latter are shown to be
disfavored by using a conservative upper bound on the $\beta \beta 0 \nu$
nuclear decay rate, if neutrinos are assumed to be Majorana particles.
Comparisons are also made with restrictions from tritium $\beta$-decay
and cosmology.
%\vspace{.10in}
\newpage

Any observation of neutrinoless nuclear double beta decay would imply
lepton nonconservation and a nonzero neutrino Majorana mass $M_{ee}$.
The latter is defined as
\begin{equation}
M_{ee} = \sum_i m_i U_{ei}^2 \ \ ,
\end{equation}
where $m_i$ is the nonnegative $i$th physical Majorana mass and
$U_{ei}$ the matrix element which mixes the electrons neutrino
$\nu_e$ with the mass eigenstate $\nu_i$. There is now a considerable amount
of flavor oscillation data from solar\cite{bahcall} and atmospheric\cite{toshito} 
neutrinos, all of which can be accommodated
within the standard picture of three neutrinos $\nu_e, \nu_\mu,
\nu_\tau$ with tiny masses.
The best fits yield two independent squared mass differences among the neutrinos:
$\Delta_S \sim 4\times10^{-5} \ eV^2$ for the solar case and $\Delta_A
\sim 3\times 10^{-3} \ eV^2$
for the atmospheric one, the favored values of the corresponding mixing angles
being $\sin^2 2\theta_S \sim 0.66$ and $\sin^2 2\theta_A \sim 1$.
The following question then emerges: how is the overall
mass scale of the neutrinos constrained?  Specifically, how does one
pin down the sum of the physical neutrino masses $\Sigma_\nu$ which controls the
neutrino component of dark matter and hence neutrino effects on
structure formation?

If we assume the neutrinos to be Majorana particles, there is a link 
between $\Sigma_\nu$ and $M_{ee}$.  This link  
has been the subject of several recent investigations
[3-8]. In particular, Barger et al\cite{barger} have given upper and lower
bounds on $\Sigma_\nu$ in terms of $M_{ee},  \Delta_A$  and $\theta_S$,
neglecting $\Delta_S$ in comparison with
$\Delta_A$.  When the
small mixing angle relevant to unobserved neutrino oscillations at the
CHOOZ reactor\cite{apollonio} is ignored, their inequalities become
particularly simple, namely
\begin{equation}
2M_{ee} + \sqrt{M_{ee}^2 \pm \Delta_A} < \Sigma_\nu <
\frac{2M_{ee}}{|\cos 2\theta_S|} +
\sqrt{ \frac{M_{ee}^2}{\cos^2 2\theta_S} \pm \Delta_A} \ \ .
\end{equation}
In eq.(2) the $+ \ (-)$ sign refers to the normal (inverted) three neutrino
mass\footnote{The mass ordering $m_1 < m_2 < m_3$ with nonnegative $m$'s
has been chosen by definition.} hierarchy $m_1 \leq m_2 <
m_3 \ (m_1 < m_2 \leq m_3)$.  The inequality $M_{ee} >
\sqrt{\Delta_A}$ is then automatically implied for the inverted
hierarchy case.  
However, such considerations
completely ignore another item of neutrino flavor oscillation
information, namely the data\cite{aguilar} from the LSND experiment. These data
can be explained by 
$\bar{\nu_\mu} \rightarrow \bar{\nu_e}$ (and $\nu_\mu \leftrightarrow
\nu_e)$ oscillations with a mass squared difference $\Delta_L$= $\cal{O}$(1)
$eV^2$ and a small mixing angle $\theta_L$= $\cal{O}$(10$^{-2}$).  This note is
addressed to a generalization of eq.(2) to include the LSND results.

A fourth light neutrino $\nu_s$, which is not electroweak active and is hence
called sterile, is needed along with $\nu_e, \nu_\mu$ and $\nu_\tau$ to 
simultaneously explain the solar, atmospheric and LSND anomalies.  
Of course, it follows from the
recent SNO\cite{ahmad} and Super-K\cite{toshito} results that the final state to which the
solar $\nu_e$ or the atmospheric $\nu_\mu$ oscillates cannot be a purely
sterile species.  On the other hand, orthogonal linear combinations of
$\nu_\tau$ and $\nu_s$ are still allowable \cite{whisnant} final states in these
oscillations.  Comprehensive analyses [12-14] have recently been made of
all current data on solar, atmospheric and LSND oscillations, together
with constraints from other accelerator and reactor data, by considering
the four neutrinos $\nu_e, \nu_\mu, \nu_\tau$ and $\nu_s$.  The
conclusion is that the four neutrino picture is not excluded, though the
required fits are not of particularly high quality.
\begin{figure}[!t] 
\centerline{\epsfysize 2.75 truein \epsfbox{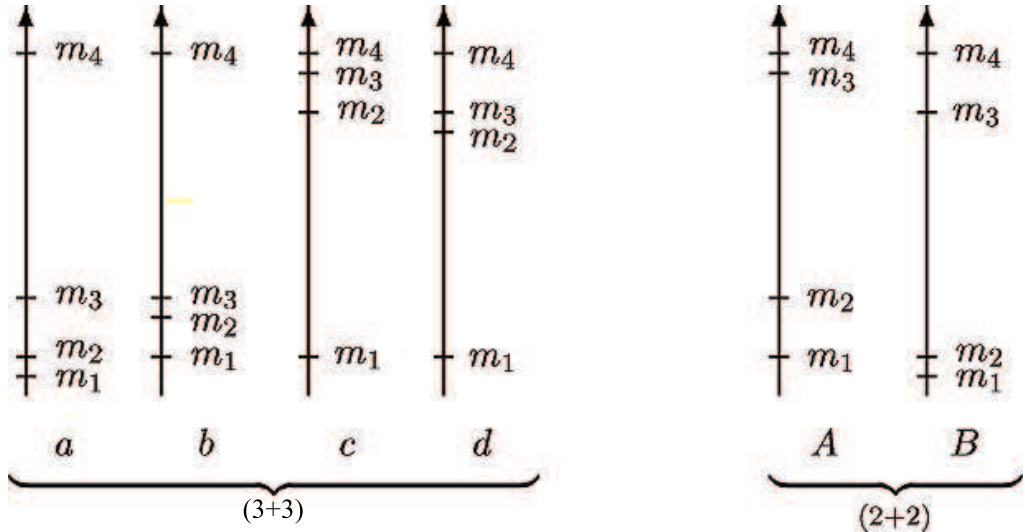}}
 \caption{Six allowed of patterns of masses, grouped into two schemes,
 for the four neutrino scenario.  Vertical separations symbolize mass
 squared differences pertinent to solar, atmospheric and LSND oscillations.}
\end{figure}

Once one considers a four neutrino scenario, with the experimental
input $\Delta_S << \Delta_A << \Delta_L$, the mass 
spectrum of the neutrinos becomes an issue of central importance.  
There are six possible 
four neutrino mass patterns, as shown in Fig. 1, that are a priori 
compatible \cite{gonzalez,maltoni,gonzalez1} with the data\footnote{Our
ordering for the physical masses is always $m_1 < m_2 < m_3 < m_4$.} 
These group into two schemes called\cite{barger1} {\bf (3+1)} and {\bf (2+2)}. 
The {\bf (3+1)} scheme, consisting of four 
possibilities  a, b, c, d (c.f. Fig.1) is characterized by three close-by
neutrino masses separated from the fourth by a gap
$\cal{O}$ ($\sqrt{\Delta_L})$.  
Here the sterile neutrino is only 
slightly mixed with the active ones. It is therefore a weak component in solar and 
atmospheric neutrino oscillations, but mainly provides a description of the LSND effect.  
In the {\bf (2+2)} scheme, comprising two possibilities A and B
(cf. Fig.1), there are two pairs of nearly degenerate states,
separated by a gap $\cal{O}$ ($\sqrt{\Delta_L})$.  In this pattern two 
orthogonal linear combinations of $\nu_s$ and $\nu_\tau$ with
comparable coefficients make up the final 
states to which the solar $\nu_e$ and the atmospheric $\nu_\mu$
oscillate.  Oscillation phenomenology alone 
cannot distinguish between different patterns within any of these
schemes.  However, a distinction does become possible
when nuclear $\beta \beta 0 \nu$ decay is taken into account \cite{bilenky}, assuming
that the neutrinos are Majorana particles.

Turning towards mixing aspects, let us define the unitary transformation
\begin{equation}
\left (
\begin{array}{c}
\nu_e \\
\nu_\mu \\
\nu_\tau \\
\nu_s
\end{array} \right )
\ = \  
\left (
\begin{array}{cccc}
U_{e1} & U_{e2}  & U_{e3}  & U_{e4}  \\
U_{\mu 1} & U_{\mu 2}  & U_{\mu 3}  & U_{\mu 4}  \\
U_{\tau 1} & U_{\tau 2}  & U_{\tau 3}  & U_{\tau 4}  \\
U_{s1} & U_{s2}  & U_{s3}  & U_{s4}  \\
\end{array} \right )
\left (
\begin{array}{c}
\nu_1 \\
\nu_2 \\
\nu_3 \\
\nu_4
\end{array} \right ) \ \ .
\end{equation}
The $4 \times 4$ matrix $U$ can be written as a product
of a $4 \times 4$ MNS 
type of a matrix $V$\cite{maki} times a
Majorana phase matrix \cite{bilenky1} diag. (1 $e^{i \alpha} \ e^{i \beta} \ e^{i \gamma})$.  
The Majorana phases $\alpha,\beta, \gamma$ make no 
contribution to neutrino oscillations but can affect $\beta \beta 0 \nu$ decay.  
The matrix $V$ in general has\cite{he} six
angles and three phases.  However, major simplifications occur  
when some experimental constraints are imposed.
We demonstrate the way $U$ is simplified in one case, namely for the pattern
{\bf (2+2)}$_B$.  The form of $U$ for other patterns can then be obtained by
interchanging some columns.

In the {\bf (2+2)$_B$} pattern (c.f. Fig.1) $\nu_e$ resides 
largely in the state $\nu_1$.
Moreover, $\nu_1$ and $\nu_2$ are the oscillating pair for
solar neutrinos and so, $\theta_{12}$, the angle 
of rotation in the 1-2
plane, can be identified with the solar neutrino 
mixing angle $\theta_S$. Any mixing between $\nu_e$ and the more massive states $\nu_3$ 
and $\nu_4$ is going to be strongly
constrained by the Bugey experiment\cite{declais} which implies that
\begin{equation}
| V_{e3} \mid^2 + \mid V_{e4} \mid^2 \ < \ 10^{-2} \ \ .
\end{equation}
We shall interpret this result to mean that, for the mass pattern ${\bf (2+2)_B}$, 
it is a good approximation to let the elements $V_{e3}$
and $V_{e4}$ be zero and replace $V_{e1}$ and $V_{e2}$ by $\cos \theta_S$ and $\sin \theta_S$
respectively.  Then the $U$ matrix of eq.(3) becomes

\begin{eqnarray}
U{_{{\bf (2+2)}}{_{B}}} \simeq \left (
\begin{array}{cccc}
\cos \theta_S     &  \sin \theta_S   & 0 & 0 \\
U_{\mu 1}     & U_{\mu 2}  & U_{\mu 3} & U_{\mu 4} \\
U_{\tau 1}     & U_{\tau 2}   & U_{\tau 3} & U_{\tau 4}
\\
U_{s 1}     & U_{s 2}  & U_{s 3} &
U_{s 4}
\end{array}\right)
\left (
\begin{array}{cccc}
1 &                  &       &  \\
  & e^{i \alpha}   &       &  \\
  &                  & e^{i \beta}    &  \\
  &                  &                  & e^{i \gamma}
\end{array} \right )
\end{eqnarray}
\begin{eqnarray*}
 & =  \left (
\begin{array}{cccc}
\cos \theta_S     &  \sin \theta_S e^{i \alpha}   & 0 & 0 \\
U_{\mu 1}     & U_{\mu 2} e^{i\alpha}   & U_{\mu 3} e^{i \beta} &
U_{\mu 4} e^{\gamma} \\
U_{\tau 1}     & U_{\tau 2} e^{i\alpha}  & U_{\tau 3 } e^{i \beta} &
U_{\tau 4} e^{i\gamma}
\\
U_{s 1}     & U_{s 2} e^{i \alpha}   & U_{s 3} e^{i \beta} &
U_{s 4} e^{i \gamma}
\end{array} \right ) \ \ .
\end{eqnarray*}

Thus the Majorana mass effective for $\beta \beta 0 \nu$ decay is given for this pattern by
\begin{equation}
M_{ee} = \mid m_1 \cos^2 \theta_S + m_2 e^{2 i \alpha} \sin^2 \theta_S
\mid \ \ .
\end{equation}
On the other hand, the nonnegative physical masses for the above pattern can
be defined, ignoring $\Delta_S$ and $\Delta_A$    
in comparison with $\Delta_L  \equiv  m^2_3 - m^2_1$, as
\begin{eqnarray}
m_1 \simeq m_2   \equiv m  \ \ ,  \\
m_3 \simeq m_4 = \sqrt{m^2 + \Delta_L} \ \ .
\end{eqnarray}
Thus we can rewrite eq.(6) as
\begin{equation}
M_{ee} \simeq \mid \cos^2\theta_S + \sin^2\theta_S e^{2 i \alpha}|m
\end{equation}
and note that in eq.(9) $m$ gets minimized (maximized) when the two terms are in phase 
(out of phase) at the value $M_{ee} \ (M_{ee}/|\cos 2
\theta_S|)$.  Since the sum of the four neutrino Majorana masses
\begin{equation}
\Sigma^{(4)}_\nu \simeq 2m +2 \sqrt{m^2 + \Delta_L}
\end{equation}
is a monotonic function of $m$, we obtain the lower and upper bounds on $\Sigma^{(4)}_\nu$
\begin{equation}
2 (M_{ee} + \sqrt{M_{ee}^2 + \Delta_L}) < \Sigma^{(4)}_\nu < 2 \left (
\frac{M_{ee}}{\mid \cos 2\theta_S \mid} + \sqrt{ \frac{M^2_{ee}}{\cos^22
\theta_S} + \Delta_L} \right )
\end{equation}
for the pattern {\bf (2+2)$_B$}.

Similar upper and lower bounds can be derived on $\Sigma_\nu^{(4)}$ 
as monotonic functions of 
$M_{ee}$ for the other five 
allowed mass patterns.  The derivation becomes very simple once it is
realized 
that one can go from one pattern
to another simply by interchanging a set of mass-eigenstate indices.  
The latter is
tantamount to interchanging the corresponding columns in the matrix $U$.  
In each case $\Sigma^{(4)}_\nu$ is a monotonic function of $m$ and the
upper bounds are obtained by putting $m_{max}= M_{ee}/|\cos 2\theta_S|$.  
However, for the lower bounds, while $m_{min}= M_{ee}$ for the patterns
{\bf (3+1)$_a$}, {\bf (3+1)$_b$} and {\bf (2+2)$_B$}, it is
Max.$(M_{ee}, \sqrt{\Delta_L})$ for {\bf (3+1)$_c$}, {\bf (3+1)$_d$}
and {\bf (2+2)$_A$}.  
We list all these
results   
in Table 1, including eq.(12) for the pattern
{\bf (2+2)$_B$} along with statements on the necessary index interchanges.  An
inspection of the entries in this table 
tells us right away that the patterns {\bf (2+2)$_A$}, {\bf (3+1)$_c$}, 
\begin{table}[!t]
\caption{Bounds on $\Sigma^{(4)}_\nu$ in six mass
patterns of the four neutrino scenario, neglecting $\Delta_{S}, \Delta_A$
in comparison with $\Delta_L$.}
%\vspace{10pt}
\begin{center}
\begin{tabular}{|c|c|c|c|}%\hline
Pattern & Interchange of mass eigenstate    & Lower bound on
$\Sigma_\nu^{(4)}$  &  Upper bound on $\Sigma_\nu^{(4)}$ \\
  & indices with respect to {\bf (2+2)$_B$}    &   & \\ \hline 
&  &  & \\
{\bf (2+2)$_B$}  &  Not necessary  & $2(M_{ee} + \sqrt{M_{ee}^2 + \Delta_L})$    
                          &  $2 \left ( \frac{M_{ee}}{|\cos 2\theta_S
                          \mid} +  \sqrt{ \frac{M_{ee}^2}{\cos^2 2\theta_S} +
                          \Delta_L} \right )$  \\ \hline
&  & & \\
{\bf (2+2)$_A$}  &  1$\leftrightarrow$3 & $2
Max.\left(\sqrt{\Delta_L}, \ M_{ee} + 
\sqrt{M_{ee}^2 - \Delta_L}\right)$    
                          &  $2 \left ( \frac{M_{ee}}{|\cos 2\theta_S|}+  
 			\sqrt{\frac{M_{ee}^2}{\cos^2 2\theta_S} -
                          \Delta_L} \right )$  \\
  & 2$\leftrightarrow$4  &  & \\ \hline
 &   &  &  \\
{\bf (3+1)$_a$}  &  Not necessary  & $3M_{ee} + \sqrt{M_{ee}^2 + \Delta_L}$    
                          &  $3 \frac{M_{ee}}{|\cos 2\theta_S|}
                           +  \sqrt{ \frac{M_{ee}^2}{\cos^2 2\theta_S} +
                          \Delta_L}$  \\ 
&   &    &    \\   \hline
{\bf (3+1)$_b$}  &   &   &  \\
  &   & $2M_{ee} + \sqrt{M_{ee}^2- \Delta_A}$   & $2
\frac{M_{ee}} {|\cos 2\theta_S|}$ + $\sqrt{\frac{M_{ee}^2}{\cos^2 2\theta_S} -
                          \Delta_A}$  \\
   &   1$\rightarrow 2\rightarrow 3\rightarrow$1 &  & \\
& & + $\sqrt{M_{ee}^2 +\Delta_L}$   & + $\sqrt{\frac{M_{ee}^2}{\cos^2 2\theta_S} +
                          \Delta_L}$  \\
  &    &    &  \\ \hline
{\bf (3+1)$_c$}  &    &   &  \\ 
  &   & $Max.(2\sqrt{\Delta_L}+ \sqrt{\Delta_L-\Delta_A}, \ 2M_{ee}$
	
&  $2 \frac{M_{ee}}{|\cos 2\theta_S|}
                           +  \sqrt{ \frac{M_{ee}^2}{\cos^2 2\theta_S} -
                          \Delta_A}$ \\
&   1$\leftrightarrow$4   &   &  \\  
&                    & + $\sqrt{M_{ee}^2-\Delta_A} + \sqrt{M_{ee}^2 -
                           \Delta_L})$  & + $\sqrt{\frac{M_{ee}^2}{\cos^2 2\theta_S} -
                          \Delta_L}$  \\
   &   &    &  \\ \hline
{\bf (3+1)$_d$}  &   &   &  \\
  &   & $Max.(2 \sqrt{\Delta_L} + \sqrt{\Delta_L+\Delta_A}, \ 2M_{ee}$ 
& $2 \frac{M_{ee}}{|\cos 2\theta_S|}$
                          +  $\sqrt{ \frac{M_{ee}^2}{\cos^2 2\theta_S}
+\Delta_A}$  \\ 
&  1$\rightarrow 2\rightarrow 3\rightarrow 4\rightarrow 1$ &   &  \\
& & + $\sqrt{M_{ee}^2 +\Delta_A} +  \sqrt{M_{ee}^2 -\Delta_L})$   & + $\sqrt{ \frac{M_{ee}^2}{\cos^2 2\theta_S} -
                          \Delta_L}$  \\ 
   &   &    &  \\ \hline
\end{tabular}
\end{center}
\end{table}
and {\bf (3+1)$_d$} are consistent only if the following inequality is satisfied:     
\begin{equation}
M_{ee}^2> \Delta_L \cos^2 2 \theta_s \ \ .
\end{equation}

Currently, the best fits\cite{gonzalez} of all oscillation data 
in the four neutrino scenario, as given in 
Table 2 of Ref.\cite{maltoni}, require $\Delta_L$ to be 1.74
$eV^2$ in the four patterns
of the {\bf (3+1)} scheme and 0.87 $eV^2$ in the two patterns of the
{\bf (2+2)} scheme.
The present experimental upper bound\footnote{A nonzero value of $M_{ee}$ in
the range 0.05 eV $< M_{ee} < 0.84$ eV has recently been claimed
\cite{klapdor1}, but there has been a strong criticism
\cite{alseth} of this alleged observation.} on $M_{ee}$ can be given
\cite{klapdor0,elliott} as $0.35 \alpha$ eV, where $\alpha$ is the uncertainty in our
knowledge of the nuclear matrix element involved in $\beta \beta 0 \nu$
decay.  It has been inferred \cite{ferruglio} from a survey of all
existing calculations that $\alpha <2.8$.   It
would therefore be safe to regard 0.98 eV as a conservative upper bound on
$M_{ee}$.  On substituting these numbers, we find that the range of
variation in the $\Sigma_\nu^{(4)}$ as a function of $M_{ee}$ is
sizeable for each of the pattern {\bf (3+1)$_a$}, {\bf (3+1)$_b$} and
{\bf (2+2)$_B$}. However, this range is found to be extremely restricted for
each of the  remaining three patterns; therefore these three
patterns, namely  {\bf (3+1)$_c$}, {\bf (3+1)$_d$} and {\bf (2+2)$_A$}
are disfavored.

We can comment on the effective mass of the electron neutrino $M_e$ which can be measured
in tritium $\beta$-decay.  The latter is given in our notation by
\footnote{The validity of this expression for the effective mass
extracted from endpoint measurements in tritium $\beta$-decay is
discussed in Ref. [4].} 
\begin{equation}
M_e = \sqrt{\sum_i m_i^2 |U_{ei}|^2} \ \ .
\end{equation}
There is an interesting inequality between $M_{ee}$ and $M_e$ which
holds in all six cases as well as the three flavor case \cite{weiler}.  It can be
expressed in two equivalent ways;
\begin{eqnarray*}
\begin{array}{lr}
M_{ee} < M_e < M_{ee}/|\cos 2\theta_S| \ \ ,
					   &   \quad \quad \quad \quad
					   \quad \quad (14a)     \\
   &   \\
M_e |\cos 2\theta_S| < M_{ee} < M_e \ \ .   &   \quad \quad \quad \quad \quad \quad  (14b)
\end{array}
\end{eqnarray*}
The content of eqs.(14) is nontrivial 
since recent solar neutrino data suggest \cite{bahcall,gonzalez} that the concerned
flavor mixing is not maximal i.e. $|\cos 2\theta_S| > 0$. The current
experimental \cite{bonn} upper bound on $M_e$ is 2.2 eV. Which of 
the above inequalities becomes interesting will depend on
whether a nonzero value of $M_{ee}$ or $M_e$ is discovered first.

\begin{figure}[!htp]
%\centerline{\epsfysize 5.0 truein
\centerline{\epsfysize 2.5 truein \epsfbox{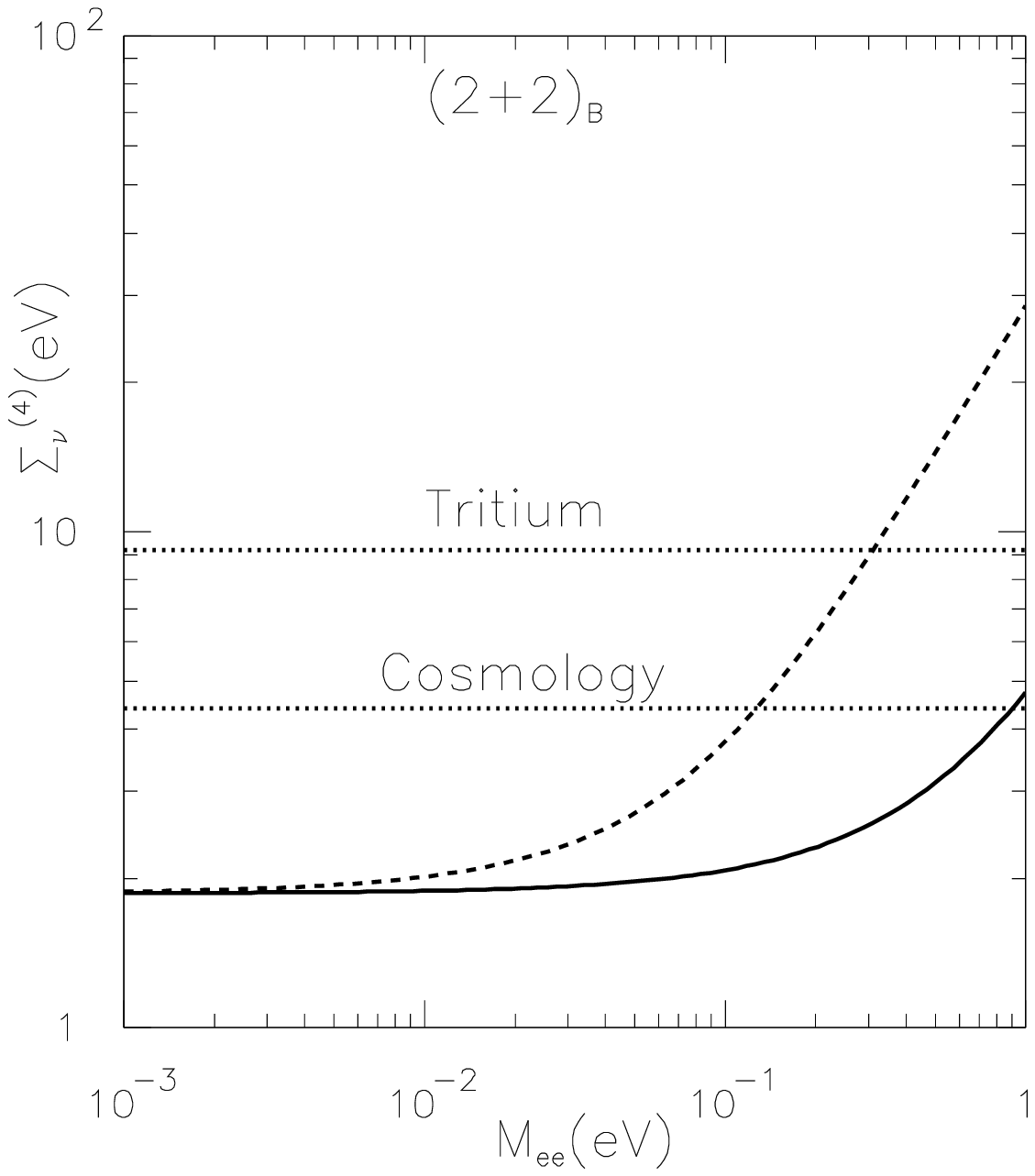}
\epsfysize 2.5 truein \epsfbox{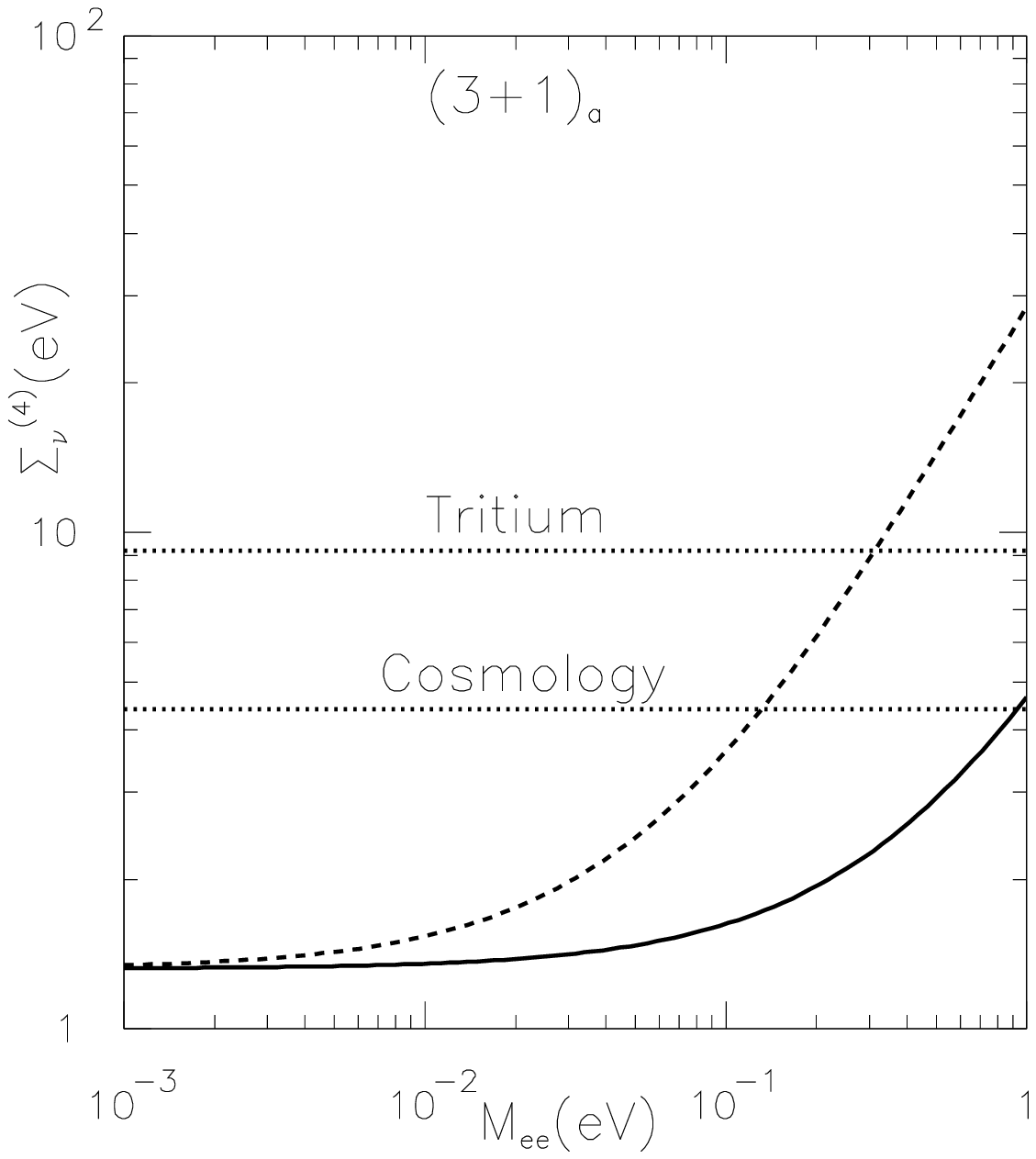}}
\centerline{\epsfysize 2.5 truein \epsfbox{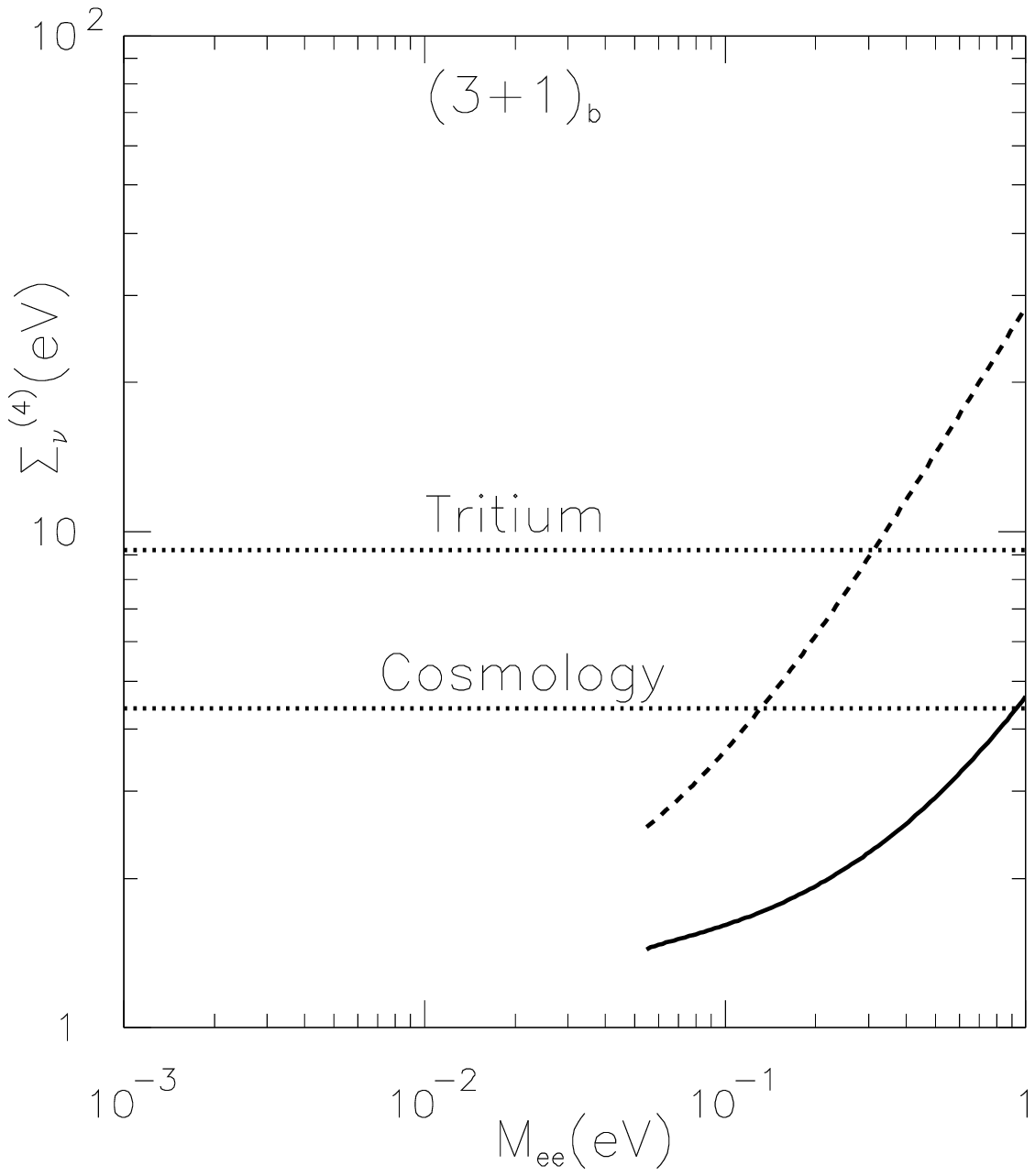}}
 \caption{Plots of the upper (dashed) and lower (bold) bounds on
$\Sigma_\nu^{(4)}$ as functions of $M_{ee}$ for the mass patterns
{\bf (2+2)$_B$}, {\bf (3+1)$_a$} and {\bf (3+1)$_b$}.  Horizontal dotted
lines show the upper bounds from tritium $\beta-$decay and cosmology.}
\label{plots}
%\end{center}
\end{figure}

A quantity of cosmological interest is the sum of neutrino masses
contributing to the hot dark matter in the Universe.  Galactic
surveys and cosmic microwave background observations bound the latter
from above by \cite{wang} by 4.2 eV.  Big Bang nucleosythesis
considerations dictate that the density of a purely sterile
neutrino species in the Universe is less\footnote{This presupposes that
the active-sterile mixing is small, specifically the effective
$\Delta m^2 \sin^4\theta < 5 \times 10^{-6} \ eV^2$, cf. Foot and Volkas
\cite{26}
and references therein.  Since this inequality is violated in the four
neutrino oscillation fits \cite{gonzalez,maltoni,gonzalez1}, we can take
the sterile density as comparable to active ones.} than that of an active
one.  But we are allowing substantial mixing between sterile and active
neutrino types.  As a result, the active density cannot significantly
exceed the sterile density.  Under the circumstances, it is not
unreasonable to treat 4.2 eV as a cosmological upper bound on
$\Sigma_\nu^{(4)}$.

We can make more precise estimates of the $\beta \beta 0 \nu$ bounds
on $\Sigma^{(4)}_\nu$.  For the LMA solution of solar neutrino
oscillations in the four neutrino scenario, we can take [13] $\sin^2 2
\theta_S < 0.98$.  Feeding in the earlier-mentioned best-fit values [13]
of $\Delta_L$, we plot the upper and lower bounds on
$\Sigma^{(4)}_\nu$ as a function of $M_{ee}$ in Fig.~\ref{plots} for the mass
patterns {\bf (2+2)$_B$}, {\bf (3+1)$_a$}, {\bf (3+1)$_c$}.  For
comparison, the upper limits from cosmology and tritium $\beta$-decay
are also shown.  A reduction in the upper limit of the allowed range of
values for $\sin^2 2 \theta_S$ in the LMA solution will tighten the
$\beta \beta 0 \nu$ bounds on $\Sigma^{(4)}_\nu$, making them more
competitive with the tritium and cosmology limits, while the next
generation of $\beta \beta 0 \nu$ experiments [25] are expected to lower
the upper bound on $M_{ee}$.  On the other hand, a significant
improvement of the cosmological bound will enable a further
discrimination among the surviving four neutrino mass patterns.

\section*{Acknowledgement}

We thank Vernon Barger for a careful reading of the paper, and Serguey
Petcov for clarifying discussions. This work is
supported in part by the U.S.D.O.E. under grant \#DE-FG 03-94ER40833.
One of us(S.P.) thanks the (Department of Energy's) Institute of Nuclear
Theory at the University of Washington for their hospitality and the
Department of Energy for partial support during the completion of this work.

\end{document}